\documentclass{edp-conf}
\usepackage{graphicx}
\usepackage{amssymb}
\def\araa{ARA\&A}%
\def\apj{ApJ}%
\def\apjl{ApJ}%
\def\aap{A\&A}%
\def\mnras{MNRAS}%
\def\nat{Nature}%
\begin{document}

\TitreGlobal{SF2A 2003}

\title{The MINE collaboration: multi-wavelength observations of microquasars with INTEGRAL, prospects with HESS and ANTARES}
\author{Fuchs, Y.}\address{Service d'Astrophysique (CNRS FRE 2591), CEA Saclay, 91191 Gif/Yvette Cedex, France}
\author{Rodriguez, J.$^1$}\address{Integral Science Data Center, Chemin d'Ecogia, 16, CH-1290 Versoix, Switzerland}
\author{Chaty, S.$^1$}
\author{Rib\'o, M.$^1$}
\author{Mirabel, F.$^1$}
\author{Dhawan,~V.}\address{National Radio Astronomy Observatory, Socorro, NM 87801, USA}
\author{Goldoni,~P.$^1$}
\author{Sizun, P.$^1$}
\author{Pooley, G.G.}\address{Mullard Radio Astronomy Observatory, Cavendish Laboratory, Cambridge CB3 0HE, UK}
\author{Kretschmar, P.$^2$}
%
\runningtitle{MINE: Multi-wavelength observations of microquasars}
\setcounter{page}{237}
\index{Fuchs, Y.}
\index{Rodriguez, J.}
\index{Chaty, S.}
\index{Rib\'o, M.}
\index{Mirabel, F.}
\index{Goldoni, P.}
\index{Sizun, P.}
\index{Pooley, G.}
\index{Kretschmar, P.}
\index{Dhawan, V.}

\maketitle
\begin{abstract} 
We present the
international collaboration MINE (Multi-$\lambda$ Integral NEtwork) aimed
at conducting multi-wavelength observations of \mbox{X-ray}
binaries and microquasars simultaneously with INTEGRAL. 
The first results on GRS\,1915+105 are encouraging and those to
come should help us to understand the physics of the accretion and
ejection phenomena around a compact object. A collaboration such as
MINE could be very useful for observing quasars and microquasars
simultaneously with HESS and later with ANTARES.
\end{abstract}
%
\section{Introduction}
	Microquasars are X-ray binaries producing relativistic jets
	and thus they appear as miniature replicas of distant quasars
	and radio-galaxies (\cite{mirabelrodriguez99}). Their emission
	spectra, variable with time, range from the radio to the
	gamma-ray wavelengths.
	We present here the first multi-wavelength campaign on
	GRS\,1915+105 involving the recently launched INTErnational
	Gamma-Ray Astrophysics Laboratory (INTEGRAL, 3\,keV--10\,MeV).
	This campaign was conducted by the MINE
	(\mbox{Multi-$\lambda$} INTEGRAL NEtwork, see
	{\sf http://elbereth.obspm.fr/$\sim$fuchs/mine.html}) international
	collaboration aimed at performing multi-wavelength
	observations of galactic X-ray binaries simultaneously with
	the INTEGRAL satellite.

\section{GRS\,1915+105}
	The microquasar GRS\,1915+105 has been extensively observed
	since this source is known to be extremely variable at all
	wavelengths (see \cite{fuchs03a} for a review).  It hosts the
	most massive known stellar mass black hole of our Galaxy with
	$M=14\pm4M_{\odot}$ (\cite{greiner01Nat}).  It was the first
	galactic source to show superluminal ejections
	(\cite{mirabelrodriguez94}) in the radio domain, which has
	enabled to give an upper limit of 11.2$\pm$0.8\,kpc to the
	distance of the source (\cite{fender99}).  In addition to these
	arcsecond scale ejections, GRS\,1915+105 sometimes produces a
	compact jet which has been resolved at milli-arcsecond scales
	in radio by \cite{dhawan00}, corresponding to a length of a few
	tens of AU.  

\begin{figure}[h]
   \centering
   \vspace*{-0.4cm}
   \includegraphics[angle=-90,width=9.5cm]{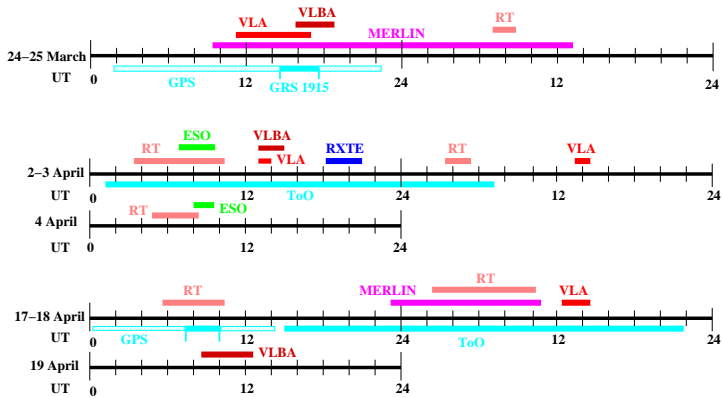}
   \vspace*{-0.3cm}
      \caption{Viewgraph of the whole observing campaign in spring
      2003, indicating the dates, time and involved observatories (GPS
      = regular Galactic Plane Survey of INTEGRAL). The
      observations discussed here are those of April 2--3.}
       \label{figcamp}
\end{figure}

	We conducted a multi-wavelength observation campaign of
	GRS\,1915+105 in spring 2003 (see Fig.~\ref{figcamp}).
	Here we focus only on April 2--3, when we obtained
	data covering the widest range of frequencies, with
	the largest number of involved instruments observing
	simultaneously with INTEGRAL (Fig.~\ref{figcamp}).
	These observations are ToO (Targets of
	Opportunity) triggered by the MINE collaboration under the
	INTEGRAL Guaranteed Time Programme 
        and related programmes on the other
        instruments.
	We thus present here
	an overview of the results of a (nearly) simultaneous campaign
	involving the Very Large Array (VLA), the Very Long Baseline
	Array (VLBA) and the Ryle Telescope (RT) in radio, the ESO New
	Technology Telescope (NTT)
	in IR, the Rossi X-ray Timing Explorer (RXTE) and INTEGRAL
	in X and $\gamma$-rays. More details can be found in \cite{fuchs03b}.

\begin{figure}[h]
   \centering
   \includegraphics[width=7cm]{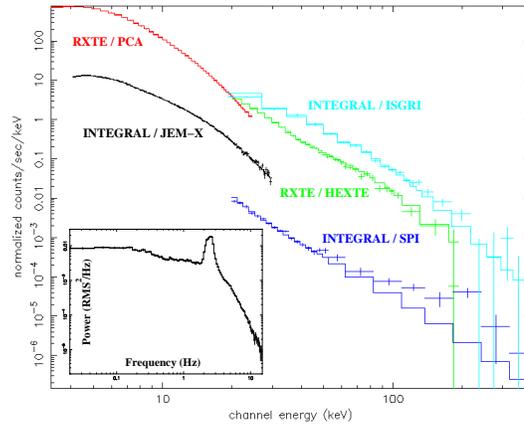}
   \vspace*{-0.5cm}
      \caption{X/$\gamma$-ray spectra of
   GRS\,1915+105 measured with RXTE (PCA \& HEXTE) and INTEGRAL
   (JEM-X, ISGRI \& SPI) on April 2, 2003. 
   Different sensitivities of the instruments lead to different levels
   of the spectra when plotted in count rates (which enables a better
   display).
   The structures at $E$$>$50\,keV in
   the SPI spectrum are instrumental background lines not adequately corrected.
   Continuous lines are the best fits 
   showing consistent photon indexes among the different instruments.
   The PCA power density spectrum (inset)
   shows a clear QPO at 2.5\,Hz.}
       \label{figspec}
\end{figure}

\begin{figure}[h]
   \centering
   \includegraphics[width=7cm]{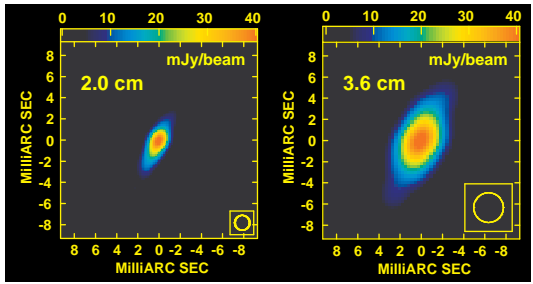}
   \vspace*{-0.3cm}
      \caption{VLBA images at 2.0 \& 3.6\,cm on April 2, 2003 showing
      the compact jet. 
	1\,mas corresponds to 12\,AU at 12\,kpc
      distance.} 
       \label{figjet}
\end{figure}

	Our
	multi-wavelength observations took place during quasi-quiet
	periods, with a slowly decaying RXTE/ASM (2--12\,keV) flux
	$\sim50$\,cts/s and an unusually high radio level 
	($>$100\,mJy at 15\,GHz with the RT).
	Such bright radio emission
	accompanied by steady X-ray emission 
	were observed on
	several past occasions in GRS\,1915+105 
	(see e.g. Fig.\,1 of \cite{muno01}).
	This state is known as the \emph{plateau}
	state (\cite{fender99,kleinwolt02} and references therein).

	This \emph{plateau} state is also called the radio loud low/hard
	X-ray state. The high energy emission of GRS\,1915+105 on
	April~2 (Fig.~\ref{figspec}) is
 consistent with this
	state, with a power law dominated spectrum (77\% at
	3--20\,keV), although always much softer (photon index
	$\Gamma$$\sim$3) than for the other BH binaries in the
	low/hard state (\cite{mcclintock03}). The INTEGRAL observations
	show that this power law spectrum extends up to 400\,keV
	without any cutoff during this \emph{plateau} state, consistent with
	the observations with CGRO/OSSE (\cite{zdziarski01}).
%
	The estimated luminosity is
	$\sim7.5\times10^{38}$\,erg\,s$^{-1}$ corresponding to
	$\sim$40\% of the Eddington luminosity for a $14M_{\odot}$
	black hole. 
	As shown in Fig.~\ref{figspec}, a very
	clear Quasi-Periodic Oscillation (QPO) at 2.5\,Hz with a 14\%
	rms level was observed in the RXTE/PCA signal, 
	which is
	consistent with the \emph{plateau} state of GRS\,1915+105.

	The VLBA high resolution images (Fig.~\ref{figjet}) show the
	presence of a compact radio jet with a $\sim$7--14\,mas length
	(85--170\,AU at 12\,kpc). This jet is very similar to the one
	observed by \cite{dhawan00} during
	the 1998 \emph{plateau} state, and is responsible for the high
	radio levels measured with the RT 
	by its optically thick synchrotron
	emission.

	The source was fairly bright in near-IR 
	with an excess of 75\% to 85\%
	in the $K_{\mathrm s}$-band flux compared to the $K$=14.5--15\,mag.
	of the K-M giant donor star of the X-ray binary (\cite{greiner01}).
	According to the spectral energy distribution, this IR excess is 
	compatible with a strong contribution from
	the synchrotron emission of the jet extending from the radio
	up to the near-IR. 
	Different components, however, contribute to the IR
	in addition to the jet,
	such as the donor-star, the external part of the
	accretion disc or a free-free emission.

\section{Conclusions and Prospects}
	Here for the first time, we observed simultaneously all the
	properties of the \emph{plateau} state of GRS\,1915+105. 
	We thus confirm the presence of a powerful compact
	radio jet, responsible for the strong steady radio emission and
	probably for a significant part of the bright near-IR
	emission, as well as a QPO (2.5\,Hz) in the X-rays 
	and a power law dominated X-ray spectrum with
	a $\Gamma$$\sim$3 photon index up to at least
	400\,keV. 
	Detailed fits of the RXTE and INTEGRAL spectra of GRS\,1915+105
	in this \emph{plateau} state, to determine for example
	whether this power law is due to an inverse Compton
	scattering of soft disc photons 
	on the base of the compact jet 
	(see e.g. \cite{fender99})
	or not, will be studied in forthcoming papers.
	In our multi-wavelength March-April campaign, the source was
	observed essentially in the \emph{plateau} state. In order to better
	understand the unusual behaviour of GRS\,1915+105, we need to
	carry out similar simultaneous broad-band campaigns during the
	other states, in particular during the sudden changes in the
	X-ray state that correspond to powerful relativistic
	ejection events.

	The latter, also observed in other microquasars such as
	XTE\,J1550--564 or V4641\,Sgr, are likely to produce very high energy
	photon and neutrino emission that could be observed with HESS
	and ANTARES, respectively.

\end{document}